# Eco-evolutionary "fitness" in 3 dimensions: absolute growth, absolute efficiency, and relative competitiveness

J. Masel

Dpt. Ecology & Evolutionary Biology, University of Arizona

1041 E Lowell St, Tucson AZ 85721, USA.

Correspondence to: masel@u.arizona.edu

Ph. +1 520 626 9888

**Running title:** relative and absolute fitness

**Abstract**

Competitions can occur on an absolute scale, to be faster or more efficient, or they can occur on a relative scale, to "beat" one's competitor in a zero-sum game. Ecological models have focused on absolute competitions, in which optima exist. Classic evolutionary models such as the Wright-Fisher model, as well as more recent models of travelling waves, have focused on purely relative competitions, in which fitness continues to increase indefinitely, without actually progressing anywhere. This manuscript proposes a new way to describe both at the same time. It begins with a revised version of $r/K$-selection theory. $r$ continues to describe maximum reproductive speed, but the new version of $K$, with a different subscript, now describes parsimoniousness in territory use, a group-selected, anti-tragedy-of-the-commons trait. A third dimension $c$ of fitness is then added to this novel system, one which is unitless and normalized, and hence capable of capturing the population genetics concept $w$ of a strictly relative, genetically-limited competitive race. MacArthur's original version of $r/K$-selection theory is shown to confound parsimoniousness $K$ with competitive ability $c$, despite the fact that available data suggests a negative correlation between the two; here they are disentangled. A rotation of the resulting three-dimensional system provides a population genetic underpinning for Grime's universal adaptive strategy theory of ruderals (selected for high $r$), stress tolerators (selected for a combination of high $r$ and high $K$), and competitors (selected for a combination of high $r$ and high $c$).

# Introduction

The flourishing field of eco-evolutionary dynamics studies how ecologically meaningful traits evolve (Kokko and López-Sepulcre, 2007; Pelletier et al., 2009; Reznick and Ghalambor, 2001; Schoener, 2011; Thompson, 1998). This forms a marked contrast to classical population genetics, which assigns strictly relative fitness $w$ values to genotypes (Fig. 1) which, by virtue of their normalization, have no absolute interpretation but travel in an unending wave without actually moving anywhere (Desai and Fisher, 2007; Mustonen and Lässig, 2010).

| Genotype | A | B |
|---|---|---|
| Relative fitness | $w_A$ | $w_B$ |
| Frequency | $p$ | $1-p$ |
| Progeny *proportional* to | $pw_A$ | $(1-p)w_B$ |
| Population mean relative fitness | $\bar{w} = pw_A + (1-p)w_B$ | |
| Absolute fitness | $w_A/\bar{w}$ | $w_B/\bar{w}$ |

**Figure 1**: Standard population genetics assignation of relative fitness to two haploid genotypes is used to calculate absolute numbers of surviving offspring.

Here I propose a new model for the fitness of a genotype. The aim is a scheme that, while remaining simple, preserves not only the distinction between relative vs. absolute competitions, but also two other core distinctions: benefits to the individual vs. the group, and reproductive speed vs. efficiency/yield. Where fitness aspects differ fundamentally in their units – as is the case for reproductive speed, maximum population density, and the unitless ability to triumph in a contest – they are given distinct representations within the scheme. Nevertheless, the high-dimensional trait space needed to describe genotypes fully is projected onto a relatively small number of parameters. The new scheme is based on synthesizing a term normalized in a $w$-like manner with reformed versions of $r$ and $K$. The focus is on "garden variety" deleterious and adaptive mutants occurring every generation in a single population, rather than on resource partitioning or other common differences between species.

Classical ecological models of the sort discussed here, which use logistic and Lotka-Volterra approaches to summarize density-dependence and/or interspecific interactions with "phenomenological" coefficients, have been largely supplanted in favor of more explicit "mechanistic" descriptions of the interactions between organisms. This trend is most pronounced in the rise of resource competition models in the tradition of Tilman (1980), Today, models deemed "phenomenological" are often viewed with suspicion (McGill and Nekola, 2010) because, compared to their "mechanistic" counterparts, they favor general description over specific predictions (McGill and Nekola, 2010; Tilman, 1980). What is more, different mechanisms can give rise to the same phenomenological model, in a manner that changes the interpretation of the phenomenological parameters (Geritz and Kisdi, 2012).

However, the detail required for most mechanistic models often restricts them, with one major aspect of fitness described in detail while other aspects are neglected. For example, the resource competition framework (Tilman, 1980), assumes no direct interactions between individuals at the same trophic level, and thus no interference competition and no sexual selection. Phenomenological models are better suited for the "bird's-eye" purpose of parsimoniously describing the interplay between major aspects of fitness. While they do not provide detail for any one aspect, they make it easier to avoid leaving important aspects out altogether. This manuscript aims for a plausible working hypothesis about how best to project high-dimensional trait space onto a simplified low-dimensional model of fitness, based on general biological considerations about the key distinctions that should be preserved: relative vs. absolute competitions, benefits to the individual vs. the group, and reproductive speed vs. efficiency/yield. Including all these distinctions between aspects of fitness is most easily achieved in the tradition of classical phenomenological models.

# Models

**MacArthur's *r*- and *K*-selection**

In a simple logistic model of population dynamics, population size $N$ obeys $dN/dt=rN(1-N/K)$. Under $r/K$-selection theory, different genotypes or species are characterized by different values of $r$ and $K$. Despite this seeming simplicity, concepts of *r*- and *K*-selection have a troubled history, with many definitions (Boyce, 1984; Parry, 1981). Some definitions focus on "fast" and "slow" life history (Jeschke et al., 2008); the relevant concepts have since been incorporated into more sophisticated life-history models of age-dependent mortality and reproduction (Reznick et al., 2002), and are not discussed here. The use of the terms $r$ and $K$ in this manuscript is instead based strictly on their behavior as parameters in the logistic equation $dN/dt=rN(1-N/K)$.

*r*-selection acts on the speed of population expansion at low density, which is a form of absolute fitness. The more controversial part of *r-K* theory is *K*-selection (Jeschke et al., 2008; Mallet, 2012), which I redefine in this manuscript. My aim is a formalism in which a *K*-adapted genotype differs only in some characteristic affecting maximum population density, e.g. via the parsimonious use of resources. In order to make a conceptually clean distinction between aspects of fitness that have different units, two genotypes that differ only in $K$ should differ in maximum population density alone, and not in low-density growth rate $r$, in resource partitioning, or in interference or other forms of competitive ability.

Canonical models of *r*- and *K*-selection (MacArthur, 1962; Roughgarden, 1971) assume that the dynamics of genotype $i$ are best described by

$$\frac{dN_i}{dt} = r_i N_i \left(1 - \frac{1}{K_i} \sum_j N_j \right) \quad (1)$$

Unfortunately, the canonical Eq. 1 is not compatible with the interpretation of a high-$K$ genotype as a parsimonious user of resources. Instead, Eq. 1 would imply that all benefits from this parsimony are directed exclusively to individuals of identical genotype. However, in the absence of spatial structure, resources left unused are normally equally available to all genotypes rather than preferentially enjoyed by the high-$K$ genotype (Chao and Levin, 1981).

To see this clearly, consider a new mutation with $K_2 > K_1$ entering a population previously fixed for genotype 1, so that $N_2$ is small and $N_1 \sim K_1$. If $K$ represents anti-tragedy-of-the-commons parsimoniousness, a small amount of resources should now be freed, giving a minute benefit to individuals of both genotypes. But in Eq. 1, genotype 1 gets no benefit, while genotype 2 gets a benefit whose size is greatest at the beginning, when genotype 2 is rare and increased $K_2$ has not yet led to an increase in the total population size $N$.

$K$ in Eq. 1 might instead be interpreted as competitive ability to dominate at high density, making the immediate, low-frequency benefit no longer a puzzle. However it is now unclear why a new, hawkish competitor genotype 2 should have a higher maximum population density than the dove genotype 1 it displaces. As shown using α-matrices in a section below, the standard Eq. 1 formulation of $K$ assumes a tight coupling between resource use parsimoniousness and competitive ability; I will propose a new formulation that disentangles these two aspects of fitness.

Perhaps the best interpretation of classical Eq. 1 $K$-selection is that this new high-$K$ mutant exploits a previously neglected resource (Levin, 1971). Innovation in resource partitioning might or might not be common in ecological speciation. However, there are reasons to believe that it is rare in adaptation. Its appearance in Lenski's experiments (Blount et al., 2008), involving the ability to exploit citrate as a result of gene duplication and associated promoter capture leading to altered regulation of the new ortholog (Blount et al., 2012), was a spectacular and newsworthy occurrence, rather than a "garden-variety" adaptation. This is despite the fact that the experimental setup, with a single species exploiting only one of two available resources, and requiring only change in the regulation of the expression of an existing gene to exploit the second, was in retrospect almost designed to make such an occurrence easy. Outside the laboratory, innovations in resource consumption may be more difficult because of competition with other species, and to the best of my knowledge, none have been documented at the genetic level. This interpretation of MacArthur's Eq. 1 $K$-selection describes an event that seems to be extremely rare in evolution.

**Alternative version of *K*-selection**

A more reasonable equation for selection on maximum population density via parsimoniousness in resource use, while holding constant other factors such as competitive abilities and resource partitioning, is described by the parameter *K* in

$$\frac{dN_i}{dt} = r_i N_i \left(1 - \sum_j \frac{N_j}{K_j}\right) \qquad (2)$$

Note the change in the subscript of *K*. In Eq. 2, $1/K_j$ can be interpreted not only as the amount of resources needed to support one individual of genotype *j*, but also as the flow of resources occupied in the process of this support, and hence unavailable to other individuals of any genotype. This failure of MacArthur's $K_i$ to describe the efficiency of conversion of biomass to offspring has occasionally been pointed out (Joshi et al., 2001) but somehow does not seem to have been pursued. As discussed in the Supplement, in order to avoid complications regarding body size, $N$ is best interpreted as biomass rather than as number of individuals.

Note that the adaptive evolution of parsimoniousness in resource use $K$, in my Eq. 2 formulation, requires group selection (MacLean, 2008). Indeed, group selection has previously been modelled via $K$ by replacing MacArthur's $\frac{\sum_j N_j}{K_i}$ with a third alternative, $\frac{\sum_j N_j}{\bar{K}}$, where $\bar{K}$ is population mean $K$ (Wilson, 1987). The Eq. 2 choice of $K$-subscript is derived by assuming that forgone resources are equally available to all members of the population, just as in a classic tragedy of the commons. In the absence of group selection in favor of higher yield, and in the absence of a pleiotropic tradeoff or genetic correlation of $K$ with $r$ or with some other as yet unspecified fitness component, two lineages with different values of $K$ in Eq. 2 have neutral evolutionary dynamics. Note however that, unlike in the classic tragedy of the commons, there is no selection for greedy resource use (low $K$) either, unless it is added as an auxiliary assumption of a tradeoff between $K$ and $r$, or between $K$ and some other yet to be determined dimension of fitness. The aim of the modeling scheme described here is first to partition fitness into conceptually distinct components, as demonstrated by their different units of reproductive speed, maximum population density, and unitless contest-winning ability, while allowing correlations and tradeoffs to be added later in the light of data.

**Adding a density independent term**

The logistic equation is subject to a known pathology; if $r$ is negative, then when $N>K$, the population grows instead of shrinking (Geritz and Kisdi, 2012; Hutchinson, 1978; Kuno,

1991; Mallet, 2012; Wilson, 1925). Even if $N<K$, when $r$ is negative then population decline is counterintuitively slowest near $K$. $r$ must therefore be constrained to be positive.

One approach to minimize this pathology is to consider both density-dependent and density-independent influences on population dynamics, via the equation

$$\frac{dN_i}{dt} = b_i N_i \left(1 - \sum_j \frac{N_j}{K'_j}\right) - \mu_i N_i \qquad (3)$$

Eq. 3 can be made equivalent to Eq. 2 by setting $r=b-\mu$ and $K=K'(b-\mu)/b$. But now, when $r=\mu-b<0$, the three negatives ($K$ as well as $N-K$ and $r$) in Eq. 2 now correctly combine to form a negative, so long as $b>0$. The combination of negative $r$ but positive $K$ in Eq. 2 does not make biological sense, and the use of Eq. 3 avoids this combination.

A simplistic interpretation of Eq. 3, given the remaining weaker constraint of $b>0$, is that birth is subject only to density-dependent selection while death is subject only to density-independent selection. A problem with this interpretation, however, is that $N>K'$ then leads to a negative number of births. Fortunately, Eq. 3 is also compatible with a less restrictive interpretation, in which the partitioning of birth and death between density-dependent and density-independent factors is not absolute, but where the correlation between birth (death) and density-(in)dependence is nevertheless strong enough to ensure that $b>0$. In this less restrictive interpretation, $b$ represents all density-dependent factors and $\mu$ represents density-independent effects on both births and deaths. $b<0$ is not permitted, while $N>K'$ is, allowing for a single negative but not a double negative, and avoiding the pathology.

Eq. 2 $K$ is the maximum sustainably achievable population size, while Eq. 3 $K'$ describes a more abstract theoretical resource limit of what the carrying capacity or equilibrium population size would hypothetically be in the absence of the constraints posed by the density-independent term $\mu$ (Berryman, 1992). More simply, when $\mu$ can be interpreted as random mortality, $K'$ rather than $K$ is the best choice to be interpreted as the parsimoniousness of resource use, i.e. the target of "$K$-selection".

Given the logistic form of Eqs. 2 and 3, the best mechanistic interpretation of the "resources" to be used with different degrees of parsimoniousness are durable resources such as territory rather than consumable resources. In this case, $K'$ gives the maximum number of nesting sites in a given space, while $K$ gives the number of nesting sites that will be occupied at steady state with random mortality; density-dependent population growth is then based on dividing resources among those individuals that have secured a site (Geritz and Kisdi, 2012; Mallet, 2012). A territorial interpretation conforms to the original intent of the "*logistique*" equation; it described the availability of farmland or "*logements*" for human cultivation (Verhulst, 1845), where each human needs a territory of size $1/K'$, and where agricultural improvements could increase $K'$.

Now consider a non-random component of density-independent birth and death, so that there is selection on genotypes that specify different values of $\mu$. In terms of the logistic Eq. 2, we have $r=b-\mu$ and $K=K'(b-\mu)/b$. In other words, selection on the density-independent term $\mu$ affects both $r$ and $K$ by an equal factor (Andrewartha and Birch, 1954; Geritz and Kisdi, 2012; Ginzburg, 1992; Mallet, 2012). Variation in $\mu$ is compatible with empirical results reviewed in the Discussion, showing that $r$ and $K$ are correlated with a slope equal to 1, results that flatly contradict the more usual assumption of an $r$-$K$ tradeoff.

**Competitive ability**

Standard population genetic $w$-selection, as formalized in deterministic replicator equations and in the stochastic Wright-Fisher and Moran models, occurs at high density with population size $N$ constant and hence presumably equal to the carrying capacity $K$. $w$-selection can be thought of as a form of competitive ability to dominate at high density, e.g. via territorial contests or attracting the best mate. $r$-selection on the "Malthusian parameter" is defined with respect to low-density phenomena, and for this reason should not be equated with $w$-selection. Differential $K'$ in my redefined system corresponds only to a form of group selection, and not to either $r$-selection or $w$-selection, both of which apply at the individual level. In yet another important difference, $w$ is normalized relative to other values of $w$ in the population, whereas $r$ and $K'$ have units of speed and population density, respectively. Both $K'$ and $w$ describe effects that are most important at high population density. But the relative competitive ability $w$ studied by standard population genetic models corresponds neither to $r$-selection nor to $K'$-selection (Clarke, 1973).

Some of the best examples of strictly relative competitions, well-described by population genetics $w$, concern sexual selection. Individuals choose the "best" mate they can find, not the "best" mate on any absolute scale. Current eco-evolutionary models that do include relative aspects of fitness nevertheless tether them to absolute fitness, e.g. linking relatively selected signals to absolute quality via a fixed informational content (van Doorn and Weissing, 2006) or tying them on an absolute scale to costs (Rankin et al., 2011). The assumption that relative competitiveness is tethered is not universally true, because evolution can find ways to "cheat", e.g. selection on males to produce a fake but persuasive signal at low cost, matched at the female end by an evolutionary race to detect cheaters. Modeling this critical "cheating" aspect of sexual selection, with Red Queen innovation races to remain in the same place, requires a strictly relative aspect of fitness, as described by population genetics $w$. The known importance of sexual selection highlights the need for any comprehensive model of evolution to include (but not be limited to) relative competitions.

In an attempt to capture $w$-like selection, I therefore introduce a third fitness parameter into the logistic equation, which I call $c$-selection. Like $w$, $c$ is normalized to the mean value of $c$ in the population. Genotypes that differ only in competitive ability $c$ show no differences on an absolute scale, e.g. in reproductive speed $r$ or in steady state population size $K$.

The importance of relative vs. absolute competitions varies with density (Kokko and Rankin, 2006). For example, at low density, organisms may "settle" for whatever rare mating opportunity comes their way, weakening the intensity of relative competitions when significant absolute population expansion opportunities are available. In contrast, high density conditions may intensify intraspecies competitive violence and/or winner-take-all mate choice dynamics, accentuating relative competitions. Similarly, territorial contests are more acute agents of selection at high density.

Let $c_i$ be the competitive ability of genotype $i$ to gain territory and/or mates under higher-density conditions, for example by winning a contest with a conspecific, in a manner aligned with the population genetics tradition of assigning genotypes a relative fitness value $w$. (In contrast, $r$ or $b$ can be seen as speedy colonization of territory and $1/K'$ as the territory needed to maintain one individual.) When we consider the behavior of the total population $\sum_i dN_i/dt$, we would like the influence of $c$ to disappear, creating a clean partition between relative competitiveness $c$ and the absolute dynamics of the population as a whole. We are also looking for an equation that describes a transition from low-density selection that depends only on $r$ to one in which $c$ is also a fitness component when at high density (Fig. 2). We would like the equation to be a generalization of Eq. 3, such that when there is only one $c$-type, we recover Eq. 3. In keeping with Eq. 3, let $D = \sum_j N_j / K'_j$ be a population-averaged measure of density that can be interpreted, for example, as the proportion of existing territory needed as a minimum to support the current population. The following equation then satisfies all of the characteristics described above:

$$\frac{dN_i}{dt} = b_i N_i \left(1 - D + \frac{c_i}{\bar{c}} D\right)(1 - D) - \mu_i N_i \qquad (4)$$

where $\bar{c}$ is population mean $c$.

In the Supplement, the addition of a $c$-term to consumable resource models, rather than to a logistic model, is described.

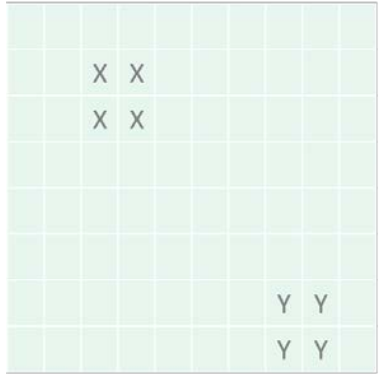 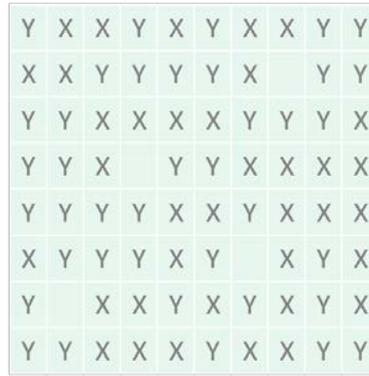

**Fig. 2.** Schematic representation of $r$- and $c$-selection. In low-density $r$-selection, all that matters is speed of reproduction to fill open space. In high density $rc$-selection with few open slots, both the rapid production of offspring as candidates to fill those spots ($r$) and the competitive ability of those candidates to do so ($c$) are important. New open spaces appear according to death rate $\mu$. Spatial structure is shown strictly for purposes of illustration; Eq. 4 describes a well-mixed population. Each individual occupies a territory of size $1/K'$ relative to total territory size of 1. $K'$ is shown here as constant such that $K'$ gives the number of territories while $K$ gives the steady state occupancy of territories.

## Comparison to α-matrix

In our system, $n$ genotypes are described by $4n$ parameters, with each genotype have a value for $b$, $K'$, $c$, and $\mu$. In an alternative approach to describing competitive ability, MacArthur's $2n$-parameter Eq. 1 is sometimes expanded by introducing an additional $n^2$ $\alpha$-parameter values (Gill, 1974; Matsuda and Abrams, 1994; Osmond and de Mazancourt, 2013) to obtain

$$\frac{dN_i}{dt} = r_i N_i \left(1 - \frac{1}{K_i} \sum_j \alpha_{ij} N_j \right) \qquad (5)$$

My Eq. 2 can be seen as a special case of Eq. 5 that is obtained by setting $\alpha_{ij}=K_i/K_j$, in the process removing $n^2$ parameters to collapse the system back to $2n$ parameters, albeit a differently defined $2n$ parameters than in MacArthur's original scheme.

Given the implicit redundancy of $K$ and $\alpha$ in Eq. 5, some authors (Kuno, 1991; Mallet, 2012) advocate abandoning $K$ altogether in favor or the equivalent formulation, with $n+n^2$ rather than $2n+n^2$ parameters, of

$$\frac{dN_i}{dt} = N_i\left(r_i - \sum_j \alpha_{ij} N_j\right) \qquad (6)$$

Eq. 6 is a very general description of genotype (or species) interactions. Unfortunately, $n+n^2$ is still a large number of parameters, making Eq. 6 difficult to use in full generality. Most models implicitly or explicitly simplify the structure of the $\alpha$-matrix in order to focus on phenomena of interest. I argue that if parsimoniousness in resource use is acknowledged to be a phenomenon of interest, then Eq. 2, rather than Eq. 1, is a better way to simplify Eq. 6 in order to isolate this effect.

Eq. 4, unlike Eq. 6, is not a competitive Lotka-Volterra equation; its density-dependence is quadratic rather than linear. But because competitive Lotka-Volterra equations of the Eq. 6 form are common in the theoretical ecology literature, I next consider a $3n$-dimensional special case of Eq. 6 whose parameters can also be interpreted as low-density reproductive speed $r$, resource use parsimoniousness $K'$, and some version of high-density competitive ability $c'$.

$$\frac{dN_i}{dt} = r_i N_i\left(1 - \frac{1}{c'_i}\sum_j \frac{c'_j N_j}{K_j}\right) \qquad (7)$$

Eq. 7 is a special case of the Eq. 6 α-matrix using only $3n$ parameters instead of $n+n^2$. In Eq. 7, $N_j/K_j$ represents the demand on territory coming from all individuals with genotype $j$. Multiplying this by $c_j'$ takes into account genotype $j$'s ability to obtain that territory. Dividing by $c_i'$ puts genotype $j$'s competitive ability into a context that is relative to the focal genotype $i$, representing the extent to which a genotype's own competitive abilities protect it from the attempts of others to take territory. The term inside the parentheses describes the extent to which genotype $i$'s growth is inhibited as a consequence of demands made by other members of the population, across all genotypes.

Unfortunately, when we consider the behavior of the total population $\sum_i dN_i/dt$, the influence of $c$ does not disappear from Eq. 7, making the absolute dynamics of the population as a whole dependent on the $c$-composition of the population. This makes Eq. 7 inferior to Eq. 4 for the purposes of a complete separation between three intuitive axes of selection, namely speed, parsimoniousness, and competitiveness. Nevertheless, the vector $c'$ in Eq. 7 captures something with the flavor of competitive ability. Eq. 4 lacks this flaw, but is not a special case of the α-matrix; values of $α$ are density-dependent rather than constant.

I introduce Eq. 7, despite this flaw, primarily in order to make clearer what is going on in MacArthur's classical Eq. 1. Specifically, by setting $c'_j=K_j$ in Eq. 7, we recover Eq. 1. In other words, MacArthur's classical formulation of "$K$-selection", when expressed in terms of our Eq. 7 model, is equivalent to the simultaneous presence of both $c'$-selection and Eq. 2 $K_j$-selection on resource use parsimoniousness alone. Instead of a clean partition, or indeed the tradeoff between $c$ and $K$ empirically supported by the literature reviewed in the Discussion below, selection on competitive ability $c'$ has the side effect of increasing the parsimoniousness of resource use $K'$ and resulting maximum population density $K$ (and vice versa). It is therefore not surprising that MacArthur's version of $K$-selection has been found to be extremely potent.

When evolution within a class of α-matrices is studied, e.g. via G-functions (Cohen et al., 1999; Vincent et al., 1993), the above discussion suggests changes to the way that classes of genotypes are defined. Since $K_j$ does not directly affect individual fitness, models using genotype classes that vary in $K_i$ (Cohen et al., 1999; Vincent et al., 1993) cannot simply be changed to ones that vary $K_j$. As discussed below, a better alternative, more consistent with both theoretical considerations and observed correlations, is to define genotype classes with respect to $\mu$, corresponding to proportional change in $r$ and $K_j$.

If the quadratic density-dependence of Eq. 4 (i.e. cubic dependence on $N$) is to be avoided, a clean partition between relative and absolute fitness can be achieved with the equation

$$\frac{dN_i}{dt} = b_i N_i \left(1 - D + \frac{c_i}{\bar{c}} D\right) - \mu_i N_i \qquad (8)$$

However, the use of $\bar{c}$ in Eq. 8 prevents compatibility with the α-matrix approach, in which each competitive inhibition coefficient is assumed to be a function only of the two species in question, independently of the overall composition of the population.

## Discussion

**Defining correlations and tradeoffs between *r*, *K*, and *c***

Evolution takes place in a high-dimensional space. The choice of $r$, $K$ and $c$ is designed to be conceptually clean, rather than to presuppose the nature of genetic correlations and phenotypic tradeoffs, i.e. the exact rotation of the space. The conceptual distinctiveness of $r$, $K$ and $c$ can be seen by the fact that they are described using different units: $r$ has units of speed, $K$ of population density, and $c$ is intrinsically unitless. An alternative approach is to build the non-constancy of $r$ and/or $K$ into models of competitive ability, based on assumptions about how phenotypic tradeoffs and costs accompany competitive strategies (Case and Gilpin, 1974; Rankin et al., 2011). The model presented here allows tradeoffs to be added later, rather than built in *a priori*. In particular, MacArthur's version of $K$-selection can be seen to have a built-in positive relationship between $c$ and $K$; here I have disentangled the two.

The nature of any tradeoffs is evolutionarily important. For example, if competitive relative fitness comes with an absolute fitness cost, then "evolutionary suicide", as a special case of the tragedy of the commons, can sometimes occur (Ferrière, 2000; Haldane, 1932; Matsuda and Abrams, 1994; Parvinen and Dieckmann, 2013; Rankin et al., 2011; Rankin and López-Sepulcre, 2005). Understanding the appropriate rotation of the space by defining the principal components within a fitness space of $r$, $K$, and $c$ (and potentially additional dimensions) will inform about the nature of tradeoffs, and hence the likelihood of such scenarios.

Principal components can be defined with respect to the distribution of new mutations, with respect to the distribution of adaptive fixation events, or with respect to standing genetic variation within or between populations. The principal components may be different in different cases. In particular, mutations and adaptations are more likely to define axes of better vs. worse, albeit as shaped by genetic correlations. Variation among populations is more likely to lie along lines of almost equally fit phenotypic tradeoffs (perhaps shaped by even stronger genetic and/or environmental correlations), although the Pareto front nature of this variation may be obscured by measurement in a single environment in which some populations perform better than others.

The principal components of variation among species may be radically different from the various kinds of within-species comparisons listed above. In ecological interactions between species, unlike evolution within a species, resource partitioning is likely to be a key player. Evolution has sometimes been seen to differ from ecology primarily in its slower timescale, albeit in a view that is now much disputed (Carroll et al., 2007; Hairston et al., 2005). Here I propose two differences between ecological (among-species) and evolutionary (within-species) processes that I believe are more substantive. First, sexual selection causes the $c$ dimension to be more important in evolution than in ecology. The second difference lies in the nature of newly introduced variation. In evolution, novel genotypes enter primarily by mutation, whereas in ecology, novel species generally enter by migration or, in special and rarer circumstances, sympatric speciation. I hypothesize that the first three principal components of the evolution of territorial populations, but not necessarily of their ecology, can be well approximated by a rotation of a space defined only by the three dimensions $r$, $K$, and $c$.

One version of this hypothesis can be seen as isomorphic with Grime's (1977) universal adaptive strategy theory, with a triangle of competitive, stress-tolerant, and colonizing traits. In Eq. 4 (after transformation from $\{b, K', \mu\}$ to $\{r, K\}$), high density selection on the product of $r$ and $c$ acts on Grime's competitive traits, while pure $r$-selection acts on Grime's colonization ability. Grime's stress-tolerance traits are arguably best related to density-independent $\mu$-selection (acting proportionately on both $r$ and $K$). Eq. 4 can therefore be viewed as a population genetic formulation of Grime's theory (Fig. 3). Using alternative terminology, reproductive speed $r$ also maps onto some concepts of "exploitation competition", $c$ onto "interference competition", and $\mu$ onto some conceptualizations of abiotic challenges.

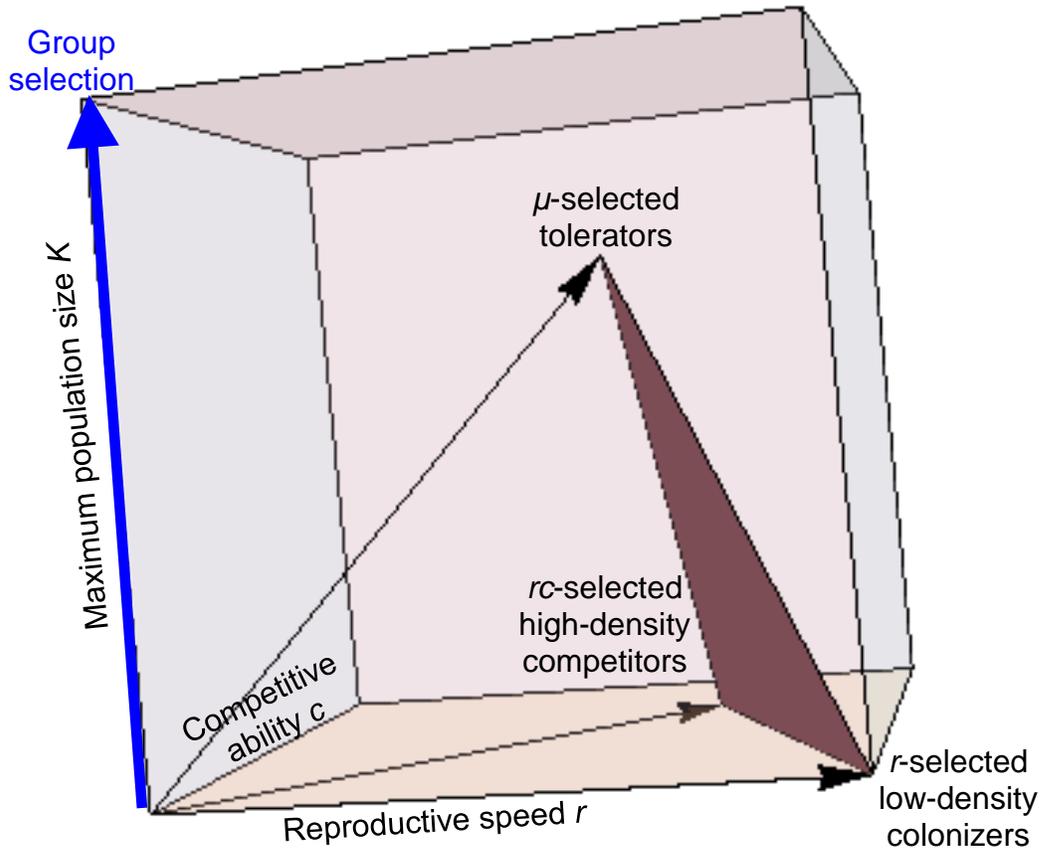

**Fig. 3:** Grime's triangle of three extreme reproductive strategies arises within the three-dimensional $rcK$ model even in the absence of tradeoffs. Selection in three environments favoring the three most extreme strategies is shown as three vectors of equal length in directions $r$, $r+K$, and $r+c$ within a cube whose axes are defined by Eq. 4. This conceptual representation of three equal vectors was chosen to illustrate that while all three forms of selection lead to increases in $r$, the increase in $r$ is greatest in the pure-$r$-selecting low-density environments encountered by colonizers. This three-dimensional representation is a projection of the four-dimensional Eq. 4. The fourth dimension, group selection $K'$, projects solely onto $K$.

**Measuring correlations and tradeoffs between $r$, $K$, and $c$**

Finding the appropriate rotation of the fitness space is a solvable empirical problem. Consider again Eq. 4, with the $\mu$ term incorporated into the main logistic to yield a three-dimensional rather than four-dimensional system. The growth rate $r$ at low density can be measured by fitting a logistic growth curve, $K$ by measuring equilibrium population size or biomass, and the product $rc$ by the outcome of pairwise, high-density competitions.

Interestingly, a number of empirical studies, in direct contradiction to MacArthur's hypothesis of an $r$-$K$ tradeoff, have found that $r$ and $K$ are positively correlated with a slope close to 1. This result has been found for different species and strains of Paramecium (Luckinbill, 1979), during group selection for high $r$ in Paramecium (Luckinbill, 1979), during selection for high $K$ in *E. coli* (Luckinbill, 1978), for a range of bacterial species-stressor combinations (Hendriks et al., 2005), for different *Nephotettix* leafhoppers species at different temperatures (Kuno, 1991; Valle et al., 1989), and among antibiotic resistant mutants of *Pseudomonas fluorescens* (Fitzsimmons et al., 2010). This points to a tight $r$-$K$ link that exists along a "better-worse" axis. Such an axis can explain why a mutation that improves fitness in one environment so often also improves it in others (Ostrowski et al., 2005).

This finding is not universal. In Lenski's experimental lines of batch *E. coli* cultures, $r$ and $K$ initially went up in tandem, but then eventually spread out across a tradeoff curve (Novak et al., 2006). When *E. coli* were instead transferred at mid-exponential, this purer version of $r$-selection did not increase yield (LaCroix et al., 2015). $K$-selection also led to a correlated decrease in $r$ in *Lactococcus lactis*, corresponding in part to a switch away from lactate production and towards acetate, formate, and ethanol metabolic end products (Bachmann et al., 2013). A tradeoff between $r$ and $K$ has been observed during the experimental evolution of haploid but not diploid *Saccharomyces cerevisiae* (Jasmin et al., 2012). Note that this latter negative *r-K* correlation was also found for variation among clones from the same yeast strain, and even in replicate studies of the same clone, pointing to an environmental cause that could confound this and other assays of closely related genotypes (Jasmin and Zeyl, 2012).

Notwithstanding the exceptions, there remains substantial evidence that a tightly coupled *r-K* axis is common, with a slope often near 1. Recall that changes in the density-independent death rate $\mu$ (as well as changes in resource use efficiency *e* in Eq. S3) affect *r* and *K* by an equal factor. The frequently observed slope of 1 between *r* and *K* therefore suggests that the density-independent death rate $\mu$ describes a first principal component across a range of circumstances, from standing variation among strains tested in the same environment, to the variety of adaptive mutations arising in the same ancestral genotype.

Competitive ability is an obvious candidate for a second principal component. The ability to emerge victorious from a high-density competition is described by the product *rc* in the transformed version of Eq. 4, combining the ability to rapidly produce offspring with the relative success of each. Competitive ability is negatively correlated with *r* (and thus, based on the discussion above, presumably also *K*) among Paramecium strains and species (Luckinbill, 1979). High density *Drosophila* populations evolve competitive ability that also trades off with low-density growth rate *r* (Mueller et al., 1991; Mueller and Ayala, 1981). This evidence for a negative correlation between relative competitive ability and absolute fitness adds empirical grounds not to entangle them, as done by MacArthur, in addition to the conceptual argument based on their having different units. In some cases, e.g. high density populations of *E. coli*, the evolution of a costly inhibitory substance may be responsible for a tradeoff between *c* and absolute fitness (Luckinbill, 1978). A principal component of *c* is a serious contender, subject to a tradeoff with $\mu$. A third principal component of $r$ alone then completes the translation of Grime's hypothesis into my scheme (Fig. 3).

Recent advances in high-throughput robotics-based experimental evolution have the potential to go beyond these literature-inspired speculations, and define the principal components more precisely in a range of different circumstances. As well as increased rigor and precision, this can increase the subtlety of the questions asked. For example, is the $r$-$K$ relationship following $r$-selection the same as the relationship following $K$-selection? For these purposes, we need to be able to vary both the nature of selection and the measurement of fitness.

Note that $K$-selection on microbial yield is group selection, which is absent from most experimental evolution setups, which lack spatial structure. But this need not be. For example, microbial metapopulations can be created in an emulsion, where many small and separated droplets of growth medium are seeded with only one cell per droplet, allowing growth up to the carrying capacity of the droplet, and then remixing and diluting to seed a new generation of droplets (Bachmann et al., 2013). This protocol selects for both increased biomass ($K$) and reduced cell size.

Microfluidics also allow for more complex protocols at the level of the droplet (Agresti et al., 2010). For example, premature colony death and/or more complex droplet "colonization" protocols introduce an additional component of $r$-selection, and the resulting balance between $r$- and $K$-selection can be altered through these parameters and/or via droplet size (Frank, 2010). Seeding droplets with more initial cells of potentially different genotypes introduces c-selection and increases $r$-selection (Frank, 2010).

If batch culture is used, transfers at mid-exponential yield *r*-selection (LaCroix et al., 2015), while less frequent transfer (Becks and Agrawal, 2013), or transfer only once a given density is reached (Yi and Dean, 2013) may yield a mix of *r*- and *c*-selection. Different conditions can yield qualitatively different outcomes; for example, when both *r*-selection and *c*-selection are present, a temporally varying environment allows coexistence rather than competitive exclusion (Dean, 2005; Yi and Dean, 2013).

Note that liquid culture, in its various forms, is likely best described by Eq. S3 resource use, while biofilm experiments may be better described by Eq. 4 territoriality. Turbidostats and chemostats are effective ways to explore Eq. S3, by allowing the control of density, and hence the study of density-dependence. Many other experiments also study density-dependence, but unwittingly. Any time that pleiotropy is studied by exposing the same genotypes to different environments, it is possible that environment-specific performance is a function not of the deliberately manipulated environmental variable, but instead of the unwittingly manipulated total population density. This is a confounding factor any time that population density depends on the environment – in other words, almost all of the time.

Understanding density-dependent fitness means assembling a panel of closely related genotypes of interest, and assaying their values of $r$, $c$, and $K$ (or of $b$, $c$, $K'$, and $\mu$, or of Eq. S3 alternatives). Assays of random mutants define the fitness axes of mutation bias. Note that different loci may consistently map to different axes (Agrawal, 2010; Laffafian et al., 2010). Assays of polymorphisms within adapted populations define selective tradeoffs. Assays of adaptive substitutions define the axes of adaptation, constrained both by mutation bias and by selective tradeoffs. Substantial evolutionary insight can be obtained by comparing these different cases.

**Usefulness as a conceptual framework**

The model presented here could in principle be used as a framework for detailed and quantitative studies of density dependence using rich datasets designed for the purpose, ideally combining chemostat, biofilm and emulsion experiments, as described above. However, where possible, data are better analyzed using a mechanistic model tailored to the specifics of that case. In those scenarios for which little is known about mechanism, however, a wide-ranging phenomenological model may be the best option.

The model proposed here is perhaps most useful in a still more basic way, as a conceptual framework to replace vestiges of the flawed $r/K$ scheme. It serves as unifying framework to ask what is and is not being studied, and to classify which conceptual category a given mechanism best fits into. There is a continuum between verbal conceptual models, formal phenomenological models, and mechanistic models. Eq. 4 falls into the middle category, and its usefulness may often lie in raising challenges to implicit assumptions that are best categorized in the verbal class, rather than substituting for the downstream explicit assumptions in the consequent mechanistic model.

Here I briefly mention two examples from the literature, one microbial and one not, where a struggle with the flaws of $r/K$ selection was evident in a published paper. I have no doubt that many more such conceptual struggles occur within research groups without making it into the final published works. Many anomalies are the result of confounding $c$ and $K$ (Joshi et al., 2001), and disappear when the three dimensional scheme of Eq. 4 is adopted as a conceptual framework. The persistence of such struggles suggests that there is something worth rescuing in the flawed $r/K$ scheme.

First, killifish populations with genetic backgrounds adapted in the wild to low density had higher $r$, equal carrying capacities, and lower competitive ability than those adapted to high density (Travis et al., 2013). In my three-dimensional scheme, the finding of equal carrying capacities in populations adapted to different densities is no longer a puzzle. This is not the case in two-dimensional $r/K$ schemes, when high-density is falsely equated with selection on carrying capacity.

Similarly in an experimental microbial system, Bull et al. (2006) noted results that could be interpreted as initial $r$-selection at low viral density, followed by "$K$-selection" later on in the experiment after density rose. However, after mentioning this $r/K$ interpretation, the authors then explicitly distanced themselves from it, primarily because of problems regarding the interpretation of $K$ (which in their case corresponds to my $c$). My model would not only provide a more suitable conceptual framework for interpreting their results, but would also suggest further experiments to measure $r$, $K$ and $c$ for isolated genotypes of interest. Metapopulation systems (Kerr et al., 2006; MacLean and Gudelj, 2006) can bring true $K$-selection into stories such as this.

**Relationship of this paper relative to other eco-evolutionary approaches**

Eco-evolutionary dynamic studies use a variety of techniques (e.g. continuous adaptive dynamics vs. discrete locus models) (Day, 2005; Fussmann et al., 2007), in order to model the evolution, on relevant timescales, of organismal properties, most commonly ecologically inspired properties such as resource use parameters. The goal is often to bring evolution in to ever more complex ecological scenarios, in order to describe the evolution of communities.

This manuscript focuses on describing fitness and adaptation within a single population, rather than expansion to complex ecological scenarios. In addition to the evolution of ecologically relevant phenotypic traits, it simultaneously models relative competitiveness traits that may have no direct ecological significance whatsoever. Nevertheless, as targets of selection, they affect the evolution of populations. I am not aware of any previous eco-evolutionary studies that allow for a strictly $w$-like term describing a purely relative fitness arms race; instead, all evolution requires changes to ecologically meaningful traits. Here I mention two examples that come closest to describing a $w$-like term. One models sexually selected markers, which of themselves are relative, but which are constrained to have constant reliability in their signal for absolute traits (van Doorn and Weissing, 2006). Another models sexual harassment, treating the benefits as strictly relative and normalized like $w$, but treating the costs as absolute (Rankin et al., 2011). While this is likely an accurate description of sexual harassment, some other modes of sexual selection, such as the faking of previously honest signals and the detection of fakes, may require only genetic innovation (or "cost of natural selection" (Haldane, 1957)) and not the payment of a true fitness cost.

When neither the benefit nor the cost of a mutation has an absolute element, evolutionary systems have no equilibrium, but instead move at a constant rate in the direction of higher $w$-fitness (Desai and Fisher, 2007; Mustonen and Lässig, 2010). This is a profound difference between population genetic models and the scenarios commonly studied using eco-evolutionary approaches today. This property of relative-fitness-based population genetics, while clearly not universally true, allows for the description of innovation without limits. Limitless innovation (as opposed to stable or metastable equilibria) seems to be a genuine property of biology, which any truly universal modeling framework should therefore include. No system describing only the uptake and metabolism of resources, but not including a normalized $c$-like term, can rise to this challenge.

Neither the common population genetics assumption that all fitness is relative, nor the common ecological assumption that all competition is indirect via resources, can be justified. A true eco-evolutionary synthesis must simultaneously account for both relative and absolute competitions. This manuscript attempts to find a satisfactory definition of fitness in terms of biologically plausible intrinsic properties of genotypes. It starts with $r/K$ selection, reforms it in an important way, and then adds a third dimension of "$c$-selection", designed to describe the relative competitions that are well described by population genetics. In other words, the simplest possible population dynamics model is synthesized with the simplest possible population genetics model that allows for unbounded fitness flux.

This manuscript defines a general model of the genotypic properties that an evolutionary model should describe, breaking them down in a novel way. The strength of Eq. 4 is its ability to describe simultaneously both an untethered arms race, and properties with absolute limits. The simplicity of Eq. 4 also provides a flexible scaffold for future extensions. This reasoning stands independently of technical choices as to which approach is taken towards modeling specific instances of change in those genotypes (Kuijper et al., 2012). In other words, it is potentially compatible with adaptive dynamics, quantitative genetics, individual-based simulation, or any other technical approach to treating the evolution of the parameters defined here.

The large literature on alternative approaches to eco-evolutionary dynamics is further reviewed in the Supplement, with connections drawn to show how Eq. 4 is unique.

**The multidimensional nature of fitness itself**

A good description of evolution must take both relative and absolute competitions into account. Clearly, fitness is sometimes relative; evolution is subject to zero-sum arms races. But not all selection can be relative, or else extinction would never occur. One modeling approach to this relative vs. absolute fitness dilemma is to assume that the truth must lie somewhere in between. This manuscript explores the possibility that the best description of evolution is not *in-between*, but *both* relative and absolute competition (Clarke, 1973). In other words, not only phenotype, but fitness itself is a multi-dimensional construct. The incommensurability of the dimensions can be seen by their different units; time for reproductive speed $r$, population density for resource use parsimoniousness $K'$, and no units for the normalizable axis of competitive ability $c$. Evolution takes place in a multidimensional fitness space defined by these and perhaps other dimensions.

The literature on incommensurable fitness components has been characterized by binary comparisons, each of which is subsumed into the higher-dimensional models proposed here. The relative fitness $w$ assigned as a property of genotypes in the Wright-Fisher model is in my $c$ dimension, while absolute fitness (with units) is in $r$ and $K'$. MacArthur's high density $K$-selection is disentangled into $c$ and $K'$ components as well as being contrasted with low density $r$-selection. Group selection is described by $K'$-selection, while $r$- and $c$-selection act on individuals. Sexual selection acts on $c$, while a very classic view of natural selection acts on $\mu$, representing a common axis of $r$ and $K$.

A basic understanding of the nature of density-dependent fitness as a function of a genotype is accessible not only conceptually, but also experimentally in the era of high-throughput experimental evolution. It is an essential building block for the unification of evolutionary and ecological theories.

**Acknowledgments:** I thank countless people for helpful discussions, including at the Banff International Research Station meeting on Mathematical Tools for Evolutionary Systems Biology, Jason Bertram for suggesting Equation 8, and the Wissenschaftskolleg zu Berlin and the National Science Foundation (DEB-1348262) for financial support.

# Supplementary Discussion for
# Eco-evolutionary "fitness" in 3 dimensions: absolute growth, absolute efficiency, and relative competitiveness

**Population genetics fitness and evolutionary rescue**

The fitness of a genotype is its contribution to the genetic material of the next generation. It can be defined either in relative terms in proportion to the contributions of other genotypes, or in absolute terms as the expected number of surviving offspring. Standard models of population genetics, such as the Wright-Fisher (Fisher, 1922; Wright, 1931) and Moran (1958) models, assign a relative fitness value $w_i$ to each genotype $i$, from which normalized absolute fitnesses are calculated (Figure 1) to ensure that the mean absolute fitness of a population is equal to 1, keeping the population size constant and equal to some assigned value $N$. Because the normalization depends on genotype frequencies, absolute fitness values are frequency-dependent in standard population genetic models (Frank, 2011; Orr, 2007), rather than being an intrinsic property of a genotype in a given environment. One obvious limitation of this standard population genetics formalism is that adaptation never leads to an improvement in the absolute flourishing of a population, nor does lack of adaptation lead to extinction. All competitive interactions are strictly relative, making population density independent of the phenotypes that evolve.

Clearly, some genotypes have more direct effects on absolute fitness and hence population size. Critical limits to adaptation, as described for example by evolutionary rescue models (Bell, 2013; Bradshaw, 1991; Bürger and Lynch, 1995; Gomulkiewicz and Holt, 1995; Gomulkiewicz and Houle, 2009; Gonzalez et al., 2013; Lynch et al., 1991; Lynch and Lande, 1993; Martin et al., 2013; Orr and Unckless, 2008) and models of the substitutional load (Ewens, 1970; Wallace, 1968; Wallace, 1975) arise only when genotypes differ in absolute fitness. Unfortunately for those seeking a simple one-dimensional population genetics model, a genotype cannot directly specify absolute fitness in the same simple way as a genotype in the Wright-Fisher model specifies relative fitness $w$. In all but the special case of fitness equal to one, this would lead to either exponential growth or exponential decline (Haldane, 1953). Evolutionary rescue models (Bradshaw, 1991; Bürger and Lynch, 1995; Gomulkiewicz and Holt, 1995; Gomulkiewicz and Houle, 2009; Gonzalez et al., 2013; Lynch et al., 1991; Lynch and Lande, 1993; Martin et al., 2013; Orr and Unckless, 2008) handle this problem by considering only the transition from decline to growth, and hence focus on the conditions for extinction vs. rescue. For more general models of the nature of population persistence after the condition for rescue has been met, more complex formulations of density dependence, such as the one proposed in this manuscript, must be assumed.

**Relationship of $r$ and $c$ to population genetics $w$**

In the Wright-Fisher model, all individuals die each generation and are replaced by new individuals chosen according to relative fitness values. In the Moran model, one individual at a time is chosen with uniform probability to die, and is immediately replaced via the reproduction of another (or the same) individual chosen with probability proportional to relative fitness. In both cases, death rates are usually constant across genotypes, while birth rates are proportional to relative fitness. In the Moran model, death and replacement are discrete processes involving single individuals, whereas in Eq. 4 they are continuous, with fractional numbers of births and deaths. A continuous model is chosen in this manuscript because it avoids unstable dynamics and the pathologies associated with $N>K'$. Note that given a continuous model, the population size $N$ in Eq. 4 is best interpreted as biomass rather than as number of individuals, and is therefore agnostic with respect to body size considerations.

By taking the limiting case of small $\mu$ in Eq. 4, we recover a scenario similar to the continuous time Moran model. As $\mu$ approaches 0, density $D$ approaches 1, the rates of birth and death approach zero, and $r$ is approximately equal to $b$. In the discrete equivalent to the continuous Eq. 4, then following each rare death at rate $\mu$, individuals compete to replace the missing individual, with probability of success proportional to $bc$ (or equivalently for infinitesimal $\mu$, $rc$). In other words, relative fitness $w$ at high density $D\rightarrow 1$ is equal to the product of two fitness components, reproductive speed $r$ and reproductive competitiveness $c$. At low density $D\rightarrow 0$, the relative fitness of genotype $i$ is $r_i$. Intermediate densities represent a linear transition between these two extreme cases (Fig. 1).

**Resource use models**

Logistic approaches have fallen out of favor in many fields of ecology, replaced by explicit resource tracking. In other words, a direct but arbitrary density-dependence term in a logistic equation has been replaced by an indirect effect of density on one or more consumable resources. A genotype's "fitness" is then its ability to acquire and use resources. This is a form of absolute fitness; like $r$ and $K$, resource use parameters are not normalized relative to competitors, they have units, and they can in principle take values low enough to cause extinction.

The standard argument is that tracking the consumption of resources is more mechanistic, with the logistic equation being an inferior phenomenological proxy (Tilman, 1980). This argument misinterprets the original intent of the "*logistique*" equation; it described the availability of farmland or "*logements*" for human cultivation, i.e. a durable rather than a consumable resource (Verhulst, 1845), where each human needs a territory of size $1/K$, and where agricultural improvements could increase $K$. Territorial competitions, including sexually selected contests for territories potentially larger than those "needed" for subsistence, are well described mechanistically by the Eq. 4 version of the logistic equation (Figure 2), where the density-independent death term $\mu$ helps distinguish between the maximum number of territories $K'$ and the steady state occupancy $K$ (Mallet, 2012). The variable weighting of the $c$ term captures the fact that at low density, territory is cheap and easy to obtain, with selection favoring rapid colonization over territorial conflict. As density increases, territorial fights intensify.

If you are studying the use of consumable resources, then the logistic equation should indeed be seen as an inferior, more phenomenological proxy. But if you are studying the occupancy of durable territory, then the logistic equation is highly mechanistic, and it is resource use equations that should be seen as the inferior, more phenomenological proxy. Relative competitions are likely more intense for territory (and for mating), so this manuscript has emphasized them and hence the logistic equation.

Indeed, in this class of ecological models, all competitive interactions take place via resource competition, i.e. exploitation competition. There are no direct interactions between individuals at the same trophic level, with all competition mediated instead via another trophic level. Interference competition might perhaps be negligible in some interspecies scenarios, especially those in which territorial considerations are less important. But if these models, traditionally used to describe competition between species, are to be extended to describe competition and hence evolution of parameters within a species, then the role of strictly relative competition in sexual and other forms of social selection should not be excluded. There is abundant evidence that sexual selection is a powerful form of competition that shapes natural populations, but which operates via direct interactions rather than indirectly via the depletion of resources. Attracting the best available mate and fighting for territory are ultimately relative operations, in which genotypes have no absolute value but are ranked in comparison to the competition. This relative, arms-race aspect of sexual selection is well described by population genetics but not by resource use models.

For this reason, as an alternative case, the integration of a $c$-term into a consumable resource use model is described briefly here. Previous theories such as Tilman's (1980) resource ratio or R* theory explicitly assume the absence of direct interactions between types, i.e. the absence of differences in $c$, such that the only form of species interaction is to quickly deplete a resource and make it unavailable for others. Under this assumption, all selection in a well-mixed population could perhaps be interpreted as $r$-selection in Eq. 4, but the analogy is phenomenological and imperfect. Here I propose an analogue of Eq. 4 in which resource density is tracked explicitly, providing a model that contains both a relative $c$ term and tracks the absolute traits of resource use.

Growth rate $r$ is the product of resource availability $R$, resource uptake rate $u$, and the efficiency $e$ with which acquired resources are used. We therefore replace $K$, $K'$ and logistic density-dependence in Eq. 2 with

$$\frac{dN_i}{dt} = Ru_i e_i N_i - \mu_i N_i$$
$$\frac{dR}{dt} = \lambda_R - \mu_R R - \sum_i Ru_i N_i$$
(S1)

where resource concentration $R$ is governed by immigration-death dynamics in addition to consumption by our focal species and where density-independent mortality $\mu$ is retained for compatibility with the logistic version of the model. Competitive resource depletion strategies appear in Eq. S1 as high values of uptake $u$ even at the expense of low efficiency $e$ in using the acquired resources. Resource use parsimoniousness described earlier by the $K'$ parameter is now described by efficiency $e$. Note that parsimoniousness was a group selected trait in Eq. 4, whereas it is a component of individual $r$-like selection in Eq. S1. This means that changes to $e$, like changes to $\mu$, have equal effects on low-density growth $r$ and equilibrium population density $K$.

Eq. S1 has a serious flaw in that the maximum growth rate $r$ is no longer an intrinsic property of a genotype but instead grows in an unbounded fashion with resource availability. However, no genotype reproduces infinitely fast even if given access to unlimited resources. Instead, an asymptotic maximum growth rate $r$ is an intrinsic property of a genotype. To capture this, we follow Monod's chemostat model (Dean, 1988; Kubitschek, 1970; Monod, 1950), and instead use Michaelis-Menten kinetics for the rate of resource uptake

$$\frac{dN_i}{dt} = \frac{R u_i e_i N_i}{R + H_i} - \mu_i N_i$$
$$\frac{dR}{dt} = \lambda_R - \mu_R R - \sum_i \frac{R u_i N_i}{R + H_i}$$
(S2)

where $H_i$ is the value of $R$ for which genotype $i$ is able to acquire resources at half its maximum rate. Maximum growth rate $r_i$ has been replaced by $u_i e_i - \mu_i$, which is now an intrinsic property of a genotype.

Solving for the equilibrium of a single genotype and its resource, the equilibrium population size $K = e_i \left( \frac{\lambda_R}{\mu_i} - \frac{\mu_R H_i}{e_i u_i - \mu_i} \right) = \frac{e_i}{\mu_i} \left( \lambda_R - \mu_R R_{eq} \right)$. This model has no exact equivalent to $K'$; the theoretical limit of $K$ in the absence of density-independent death $\mu_i$ now goes to infinity. In the logistic model, $K'$ was useful for giving us a natural sense of what "high density" means. In the resource tracking version of the model, this needs to be set arbitrarily. In Eq. 4, the $r$-$c$ transition was a function of $D = \sum_j N_j / K'_j$. Now we make $D = \sum_j N_j / N'$, where the value of $N'$ defines the meaning of "high density", to obtain

$$\frac{dN_i}{dt} = \frac{R u_i e_i N_i}{R + H_i} \left( 1 - D + \frac{c_i}{\bar{c}} D \right) - \mu_i N_i$$
$$\frac{dR}{dt} = \lambda_R - \mu_R R - \sum_i \frac{R u_i N_i}{R + H_i} \left( 1 - D + \frac{c_i}{\bar{c}} D \right)$$
(S3)

where $\frac{R u_i e_i}{R + H_i}$ in Eq. S3 serves a similar role to $b$ in Eq. 4, and where $\bar{c}$ is population mean $c$.

Note that when $D>1$, differences in $c$ outweigh differences in $r$, a situation that can be made to arise in a more pronounced fashion for Eq. S3 than for Eq. 4 by setting $N'$ to be small.

An advantage of Eq. 4 is that each genotype is described by only 4 parameters ($b$, $K'$, $\mu$, and $c$), which can sometimes be further collapsed into 3 ($r$, $K$ and $c$). If consumable resource use rather than territory occupancy is to be studied, then Eq. S3 adds realism, at the cost of requiring 5 parameters per genotype ($u$, $e$, $H$, $\mu$, and $c$) plus one more ($N'$) for the species as a whole and two more ($\lambda_R$ and $\mu_R$) to describe their common resource. If more than one consumable resource is to be studied, even more parameters are required to extend Eq. S3 to cover both multiple resources and the possibility of direct interference competition, sexual selection, or other direct interactions within a trophic level. If territorial occupancy is the primary phenomenon of interest, then Eq. 4 is superior both in terms of mechanistic underpinnings and in terms of simplicity.

**Contest and scramble**

The contrast between competitive ability $c$ and population growth speed $r=b-\mu$ partially overlaps with the distinction between "contest" and "scramble" competitions (Nicholson, 1954). Contest, like $c$, is a relative competition, while scramble, like $r$, is an absolute competition. However, the contest-scramble distinction also refers to situations in which a minimum threshold of durable territory or consumable resources is required for reproductive success; in scramble competitions, territory/resource acquisition is potentially below-threshold, whereas contest describes competitions in which the winner is guaranteed above-threshold territory/resources. This threshold-based distinction is not present in my continuous-time continuous-$N$ growth model.

**Adding more dimensions**

The range of behaviors open to a single density-dependent population with a fixed genotype can be described using only two dimensions, $r$ and $K$ (or equivalently but arguably more usefully, density-independent $r$ and density-dependent $r/K$ (Kuno, 1991; Mallet, 2012)). Interactions between multiple genotypes can greatly increase the dimensionality of the system. In this work, for maximum simplicity while still meeting my goals, I introduce only a single additional dimension $c$, bringing the total to three.

The Eq. 4 model is extensible to use more parameters to describe more phenomena, while still remaining below the $n+n^2$ parameter ceiling given by the Eq. 6 α-matrix model. The simplifying $3n$ choice for Eqs. 4 and 7 is driven by the intent to study directional selection within a single species. For ecological interactions between different species, as opposed to evolutionary interactions between different genotypes, terms may need to be added to describe resource partitioning, and resulting disruptive selection, speciation and coexistence. Co-evolutionary arms races between interference and resistance mechanisms can create nontransitive interactions, greatly complicating the $\alpha$ matrix, and demanding the explicit modeling of tradeoffs specific to the mechanism of interference that is assumed. From the evolutionary perspective, another possible axis of selection is that on variance in the number of offspring (Gillespie, 1974; Lambert, 2006; Shpak and Proulx, 2007), which affects the establishment probability of a new beneficial mutation. There are also finer distinctions within the $K'$ term (Van Dyken and Wade, 2012).

I do not deny the importance of any of these phenomena, although note that highly complex $\alpha$ matrices are more important in interspecies ecology contexts. In experimental evolution, most invading mutants either sweep to fixation or are outcompeted via clonal interference (Levy et al., 2015). My model is designed for these simple and "typical" adaptive sweeps, where mutations are normally simply good or bad, rather than altering the delicate balance between competing goals. Adaptive mutants leading to coexistence do occur, but are rare enough to be highly remarkable (Blount et al., 2008).

My intention is to propose a conceptual breakdown that is rich enough to be interesting on its own, and also useful as a starting point against which still more complex scenarios can be compared. I focus on the benefits of switching from one or two fitness dimensions to three, in the process unifying population genetics with a simple form of density-dependence. I hope that this work will provide a firm basis for extensions to even more dimensions.